\documentclass[a4paper,twocolumn]{article}
\usepackage{cite}
\usepackage{amsmath,amssymb,amsfonts}
\usepackage{algorithmic}
\usepackage[dvipdfmx]{graphicx}
\usepackage{textcomp}
\usepackage{xcolor}
\usepackage{rotating}

\def\BibTeX{{\rm B\kern-.05em{\sc i\kern-.025em b}\kern-.08em
    T\kern-.1667em\lower.7ex\hbox{E}\kern-.125emX}}

\def\xskip{\hskip 7pt plus 3pt minus 4pt}
\newdimen\algindent
\newif\ifitempar \itempartrue 
\def\algindentset#1{\setbox0\hbox{{\bf #1.\kern.25em}}\algindent=\wd0\relax}
\def\algbegin #1 #2{\algindentset{#21}\alg #1 #2} 
\def\aalgbegin #1 #2{\algindentset{#211}\alg #1 #2} 
\def\alg#1(#2). {\medbreak 
  \noindent{\bf#1}({\it#2\/}).\xskip\ignorespaces}
\def\algstep#1.{\ifitempar\smallskip\noindent\else\itempartrue
  \hskip-\parindent\fi
  \hbox to\algindent{\bf\hfil #1.\kern.25em}%
  \hangindent=\algindent\hangafter=1\ignorespaces}
\def\algsubstep#1.{\ifitempar\smallskip\noindent\else\itempartrue
  \hskip-\parindent\fi
  \hbox to\algindent{\bf\hfil #1.\kern.35em}%
  \hangindent=\algindent\hangafter=1\ignorespaces}
  
\usepackage{comment}

\begin{document}

\title{Finding Critical Nodes in Interdependent Networks with SAT and ILP Solvers
}

\author{Kyozo Hida and Tatsuhiro Tsuchiya \\ Osaka University}


\newtheorem{theorem}{Theorem}

\maketitle
\footnotetext{This work has been submitted to the IEEE for possible publication. Copyright may be transferred without notice, after which this version may no longer be accessible.}

\begin{abstract}
Infrastructure systems, such as power systems, often experience cascading failures. 
Modeling an infrastructure system as a collection of interdependent networks has recently received attention as a way to explain cascading failures. 
In this study, we propose an approach to find the set of critical nodes in an interdependent network. 
For an integer $k$, we say that a set of $k$ nodes is critical if the initial failures of these $k$ nodes result in 
the most severe cascading failure among all sets of $k$ nodes. 
This approach adopts the seminal model of interdependent networks proposed by Buldyrev et al., 
in which new link failures occur in a network if the connectivity is lost in the paired network. 
The proposed approach consists of two phases. 
In the first phase, the maximum number of failure propagation stages is computed by repeatedly solving the Boolean satisfiability problem. 
This number is then used in the second phase, where the set of critical nodes is computed using integer linear programming. 
The results of applying this approach to a variety of problem instances demonstrate that the approach is feasible for up to at least 30 nodes and can be used as the baseline to compare the performance of heuristic solutions. 
\end{abstract}


\section{Introduction}

In power systems, initial failures of a small number of components are known to sometimes lead to 
cascading failures, which results in catastrophic consequences. 
Many theoretical models have been proposed to explain this phenomenon\cite{SUN2005,Laprie2007,Baldick2009,Dong2016}. 
Recently, models that capture power systems as a set of interdependent networks have received a great deal of attention.  
This line of work was pioneered by the seminal work of Buldyrev, Parshani, Paul, Stanley, and Havlin~\cite{buldyrev:nature:2010}.
Their model represents a power system as a pair of interdependent networks: a power-line and communication network. 
The initial failures of nodes lead to the failures of their incident links in both networks. 
Link failures cause some nodes to be disconnected in one network, leading to further link failures in another network. 
In this manner, failures propagate stage-by-stage and eventually terminate, eventually resulting in a set of connected components. 

By adopting this well-known model, we can address the problem of identifying the most vulnerable parts of a system.
Specifically, we consider the set of `critical' nodes. 
We say that, for an integer $k$, the set of $k$ nodes is critical if the initial failures of the $k$ nodes lead to the most severe consequence among any sets of $k$ nodes.
Critical node detection has been extensively studied using various models in various ways. 
The proposed approach is different from that of previous studies in that 
it can determine the optimal solution under the seminal model of \cite{buldyrev:nature:2010}. 


In this study, we provide a formal proof that the critical node detection problem is NP-hard. 
This implies that no polynomial-time algorithms exist for the problem (unless P = NP, which is unlikely to be the case). 
Therefore, the proposed approach is intended to be capable of finding the optimal solution for moderate-size problem instances in feasible time. 


Our approach makes full use of mathematical programming software,  
a Boolean satisfiability (SAT) solver and an integer linear programming (ILP) solver. 
The proposed approach operates in two phases. 
In the first phase, a SAT solver is repeatedly used to obtain the maximum number of failure propagation stages. 
This number is then utilized to create an ILP problem that represents the critical-node detection problem. 
In the second phase, a set of critical nodes is obtained by solving the ILP problem. 

We implemented the proposed approach and two non-exact heuristic methods. 
By applying the proposed approach and heuristic methods to several problem instances,  
we examine the problem sizes that the proposed approach can handle as well as how good the optimal solution is compared to 
solutions obtained using the heuristic methods.  

The remainder of this paper is organized as follows. 
In Section~\ref{sec:related}, a concise survey of related work is provided. 
In Section~\ref{sec:model}, an interdependent network model is described and the problem of interest is defined, namely, the critical node detection problem. 
We also outline a proof that the problem is NP-hard. 
In Sections~\ref{sec:overview}, \ref{sec:phase1}, and \ref{sec:phase2},
the proposed approach to this problem is described.
In Section~\ref{sec:overview}, we present an overview of this approach and state that it consists of two phases. 
In Section~\ref{sec:phase1}, we describe the first phase of the approach, in which the maximum number of failure-propagating stages is computed using an SAT solver, 
and, in 
Section~\ref{sec:phase2}, we describe the second phase, where a set of critical nodes is obtained using the ILP. 
We present the experimental results in Section~\ref{sec:experiment}.
Finally, Section~\ref{sec:conclusion} concludes this paper. 

\section{Related Work}\label{sec:related}


The reliability of critical infrastructure systems is of great social interest and thus has been studied extensively. 
Some kinds of critical infrastructure systems, typically power systems, exhibit complex failure behaviors 
because they are composed of several elements that are mutually dependent on each other. 
For example, the failure of a very small fraction of a system can lead to catastrophic consequences. 

Much research has been conducted to explain such failure behaviors in infrastructure systems. 
Many recent studies have focused on \emph{interdependent network models}, which treat  
a system as a set of interdependent networks. 
Typically, a power system is represented by a combination of power and communication networks. 
Here, we consider the seminal model proposed by Buldyrev, Parshani, Paul, Stanley, and Havlin~\cite{buldyrev:nature:2010}.
In fact, \cite{buldyrev:nature:2010} introduced two models, however one is only informally described, though. 
In this study, we refer to the informally described model the \emph{simplified model} to distinguish it from the model 
formally proposed in~\cite{buldyrev:nature:2010}. 
Our study is based on the latter model. 

In the formally defined model, failures are modeled as the removals of links.  
A link in one network is removed if the end nodes of the link are disconnected in another network. 
By contrast, the simplified model assumes that every node in a network will immediately fail  if 
its corresponding node in the other network does not belong to the largest connected component. 
The simplified model can make analysis easier, but ignores the behaviors in all but the largest connected component, 
even in the early stages of failure propagation.

One problem with the simplified model was pointed out by, for example, \cite{Huang2015small}, 
The authors of \cite{Huang2015small} claimed that it is incompatible with many engineering reality 
and proposed an intermediate model where nodes can survive if they belong to a connected component whose size 
exceeds a predefined threshold. 


\emph{Critical nodes} are those whose failures or removals affect the system the most.  
Identifying critical nodes is of practical interest because it is useful for the defense and protection of a system. 
Nguyen et al.~\cite{Nguyen2013} addressed the critical node detection problem using the simplified model;
critical nodes were defined as a set of nodes whose initial failures minimized the size of the remaining connected component. 
They proved that the problem is NP-hard, and proposed heuristics algorithms to quickly find 
solutions that were not necessarily optimal, but were often satisfactory. 
In~\cite{Critical2}, mixed-integer linear programming (MILP) formulations were proposed to accurately determine critical nodes 
in the simplified model and its variants. 
MILP and integer linear programming (ILP) have been used for critical node detection  under various network models (e.g., \cite{Critical2,SANTOS2018325,FARAMONDI20183}).

In \cite{Sen2014,Banerjee2017}, instead of network topologies, Boolean formulas were used to represent interdependent networks. 
The authors of these studies call this representation model the \emph{implicative interdependency model} (IIM). 
For example, an entity survives if $a$ and $b$ both survive or $c$ survives, then 
this dependency relation is represented as $ab + c$, where $+$ denotes the logical `or'. 
The authors used ILP to solve the problem of identifying the most important nodes with various criteria. 
Additionally, \cite{Sen2014} addressed the problem of identifying the $k$ most vulnerable nodes, which is a form of the critical node detection problem.  
A common feature of these studies and ours is that stage-by-stage cascading failures are encoded in the ILP problem. 
Obtaining an IIM from the graph topologies of interdependent networks is theoretically possible, but it is not feasible in practice, as the size of the IIM  can very easily and rapidly increase. 
This is because, in the model of \cite{buldyrev:nature:2010}, each link depends on the paths connecting its two end nodes in the other network, and the number of paths increases exponentially as the network size increases. 

In addition to the underlying models, there is another clear difference between our study and the previous studies that use ILP: 
our approach computes the maximum number of possible failure propagation stages before solving the ILP. 
Using this number, the size of the ILP problem can be reduced because the ILP formulation can avoid representing stages where no failures occur. 
 In contrast, the previous studies have used trivial bounds such as 
 $(\text{the number of entities}) - (\text{initial failures})$
 to determine the number of stages to be analyzed. 

The use of a Boolean SAT solver for counting the propagation stages was proposed in \cite{Hanada2019} for the first time. 
Hence, this part of our approach can be regarded as a refinement of previous work.
In \cite{Hanada2019}, the size of the Boolean encoding depended on the paths between all pairs of nodes. 
As previously discussed, such paths increase exponentially in number and, hence the previous approach is only applicable 
to small  sparse networks. 
In contrast, the size of our proposed SAT encoding depends on the numbers of nodes and links but not paths, and is polynomial with respect to the size of the given networks. 



In addition to the models discussed above, new interdependent network models continue to be developed. 
Such models include \cite{Yagan2012,Huang2013,Huang2015,Huang2015small,Chen2014,Cai2015,SANTOS2018325}. 

\section{Model, problem, and complexity}\label{sec:model}
The model assumed in this study is that proposed by~\cite{buldyrev:nature:2010}. 
We assume that a system consists of two interdependent networks, Network~A and Network~B, which  share a set of nodes: 
$V= \{1, 2, \dots, n\}$, where $n$ denotes the number of nodes, 
and $U = \{\{i, j\} \mid i, j \in V, i <  j \}$. 
The initial topologies of these networks are represented as undirected graphs $G_A = (V, E_A)$ and 
$G_B = (V, E_B)$, where $V$ is a set of nodes defined above, and $E_A \subseteq U$ and $E_B \subseteq U$ are sets of links (edges).  
The link between nodes $i$ and $j$ is denoted by $\{i, j\}$.  

\begin{figure*}
    \centering
    \includegraphics[scale=0.5]{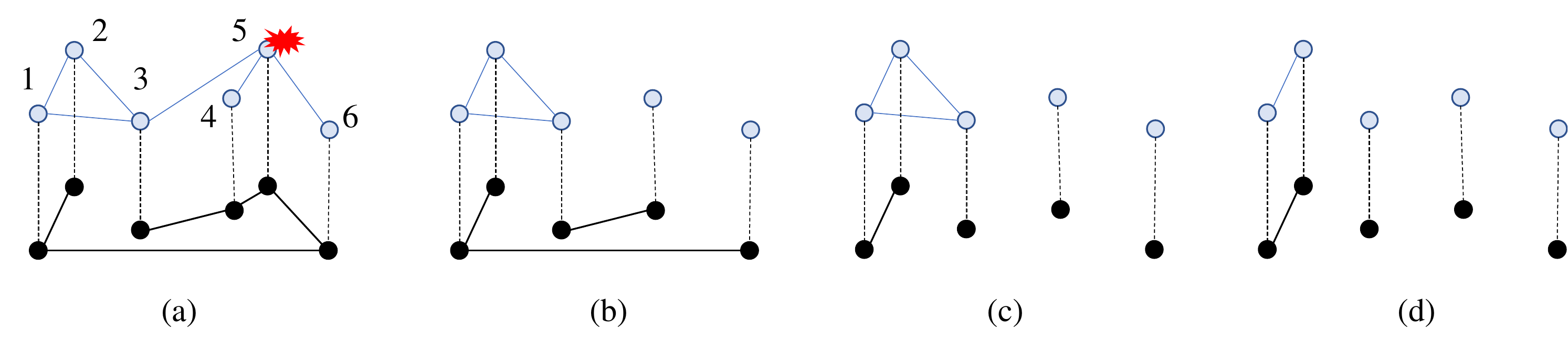}
    \caption{Failure propagation. The system consists of Networks~A (top) and~B (bottom). (a) Node~5 is attacked. 
    (b) The links incident to node~5 fail in the first stage. (c) In Network~B, the link between nodes~1 and~6 fails in the second stage 
    because they are disconnected in Network~A. The link between 3 and 4 nodes also fails. 
    (d) The link between nodes~2 and~3 in Network~A fail in the third stage.}
    
    \label{fig:example}
\end{figure*}

The cascading failures are modeled as follows. 
Failures are triggered by the initial failure of nodes. 
The nodes subject to the initial failures are called \emph{attack nodes}. 
Subsequently, the failures propagate stage by stage. 
In Stage~1, link $\{i,j\}$ is removed from the networks if either $i$, $j$ or both are attack nodes. 
The removals of links may cause some node pairs to be disconnected, thereby dividing the networks into multiple connected components. 
In even-numbered stages (Stages~2, 4, ...), the failures in Network~A propagates to Network~B as follows. 
If nodes $i$ and $j$ are disconnected in Network~A
at the beginning of the stage, link $\{i,j\}$ in Network~B is removed. 
Similarly, in odd-numbered stages (Stages~3, 5, ...), link $\{i,j\}$ in Network~A is removed 
if $i$ and $j$ are disconnected in Network~B at the beginning of the stage. 

In this model, cascading failures eventually stop at a certain stage, 
resulting in the same set of \emph{connected components} in both networks.
A connected component of an undirected graph is a maximal connected subgraph of that graph. 
The size of a connected component is defined as the number of nodes in that component. 
In addition, we say that the connected components remaining in the networks are \emph{failure-induced}. 

Fig.~\ref{fig:example} shows an example of a system modeled as a pair of interdependent networks.
The initial failure of node~5 (Fig.~\ref{fig:example}(a)) causes all the incident links to fail (Fig.~\ref{fig:example}(b)) in Stage~1. 
In Stage~2, the two links in the network at the bottom fail because their end nodes are disconnected in the 
 network above (see Fig.~\ref{fig:example}(c)). In Stage~3, the two links of the above network fail. Then, the failure 
propagation stops, resulting in four failure-induced connected components.

We now define the critical node-detection problem for the model. 
As in~\cite{buldyrev:nature:2010}, we measure the severity of cascading failures based on the size of the largest failure-induced connected component.  
Formally, the metric is given by
\begin{equation}
    f(I) := \max_{cc \in FICC(I)}  \{size(cc)\}
\end{equation}
where $size(cc)$ is the size of the connected component $cc$, 
and $FICC(I)$ is the set of failure-induced connected components when $I$ is the set of attack nodes.  
Accordingly, the problem is defined as follows.

\begin{quote}
Given $G_A$, $G_B$, and an integer $k \ (> 0)$, find the set $I$ of $k$ nodes such that $f(I)$ is minimized.
\end{quote}



In the remainder of this section, we provide proof of the NP-hardness of the critical node detection problem.
Although this proof is similar to that of the NP-hardness proof in~\cite{Nguyen2013}
for the simplified model, there are some technical differences. 
In particular, the proof in \cite{Nguyen2013} is based on the reduction from the maximum independent set problem, 
whereas our proof uses the reduction from the vertex cover problem because this simplifies the proof. 

Our NP-hardness argument follows a standard method:  
we define the decision version of the critical detection problem and prove its NP-completeness.  
The decision version of the problem is as follows.

\begin{quote}
Given $G_A$, $G_B$, and two integers $k$ $(> 0)$ and $z$ $(> 0)$, decide whether there is a set $I$ of $k$ nodes 
such that $f(I) \leq z$. 
\end{quote}

\begin{theorem}
The decision version of the critical-node detection problem is NP-complete. 
\end{theorem}

\textit{Proof}: 
First, we show that the problem is in NP. 
Suppose that a set of $k$ attack nodes is given. 
The remaining links of the networks in each stage are computed in polynomial time with respect to $n$  
because the existence of the paths between all pairs of nodes can be determined in polynomial time~\cite{Warshall}. 
Failure propagation continues for a maximum of $n$ stages; thus, it is possible to check  
if the maximum connected component resulting from $k$ attack nodes is of size $z$ or less in polynomial time. 

Next, we show that the critical node detection problem is at least as difficult as NP-complete problems. 
This is achieved by showing that there is a polynomial-time reduction from the vertex cover problem, a well-known NP-complete problem, to the critical node detection problem. 
The vertex cover problem refers to the problem of deciding, for a given $t$ and undirected graph $G$, whether 
a \emph{node cover} of size~$t$ exists. 
The node cover of a graph is a set of nodes such that all links are incident to at least 
one of the nodes. 
The size of the node cover is the number of nodes in the cover. 


Consider transforming an arbitrary instance of the vertex cover problem 
into an instance of the critical node detection problem, as follows.   
Given $G$ and $t$, we construct the latter instance such that 
$G_A = G$, $G_B$ is a complete graph, $k=t$, and $z=1$. 
Clearly, this transformation can be performed in polynomial time. 

Next, we show that this transformation is a reduction from the vertex cover problem to the critical node detection problem; the the answers to both instances always coincide.
First, suppose that $G$ has a node cover of size $t$. 
In this case, if we select the node cover as the set of $k$ attack nodes, all failure-induced connect components will be of size one ($=z$). 
This is because at least one end node of any link in $G_A$ is an attack node.  
Next, suppose there exists no node cover of size $t$. 
Regardless of how $k \ (=t)$ attack nodes are chosen, there is at least one link connecting two nodes that are not attack nodes in Network~A. 
Because Network~B is initially a complete graph, the link connecting the two nodes also remains in Network~B. 
Hence, the two nodes remain connected in both networks at a later stage, 
which results in a failure-induced connected component of size $> 1$ $(= z)$.

Because the decision version of the critical node detection problem is in NP and
a polynomial time reduction from an NP-complete problem exists, it is an NP-complete problem. 
$\square$

The NP-hardness of the critical node detection problem implies that no 
polynomial time algorithm exists for the problem unless P = NP, 
meaning that no exact algorithms can avoid super-polynomial (usually exponential) time complexity. 
Therefore, our approach aims to solve moderate-sized problem instances within a feasible time. 

\section{Overview of the proposed approach}\label{sec:overview}

The proposed approach consists of two phases:
\begin{description}
\item[1] Compute the maximum failure propagating stages by solving a series of SAT problems;  
\item[2] Find the set of critical nodes by solving an integer linear programming (ILP) problem. 
\end{description}


In effect, Phase~1 serves as a preprocessing step for phase~2.  
However, the basic idea used for problem formulation is common to both phases; 
the SAT and ILP formulations both encode the failure propagation in stages. 



In the formulation of the failure propagation, we use 
the idea underlying \emph{Warshall's algorithm}~\cite{Warshall} to compute the node connectivity.  
Warshall's algorithm is a classic method for computing transitive closures in directed graphs. 
When a graph is undirected, the algorithm can be used to compute the connectivity of any node pair. 
The undirected version of Warshall's algorithm is as follows: 
The input is an undirected graph $G$. Let $n$ be the number of nodes of $G$. 
The algorithm uses a collection of Boolean variables:  
$r_{i,j}^{(k)}$ $(1\leq i < j \leq n, 0 \leq k \leq n)$. 
For simplicity, $r_{i,j}^{(k)}$ is also denoted as $r_{j,i}^{(k)}$.  
The true and false values are denoted by 1 and 0, respectively. 
If $r_{i,j}^{(0)} = 1$ $(i\neq j)$, there is a link between $i$ and $j$ in $G$. 
For $k \geq 1$, $r_{i,j}^{(k)} = 1$, then a path exists between $i$ and $j$ that only passes through 
nodes $1, 2, 3,\dots, k$. 
Note that such a path exists if and only if 1) there is already a path between $i$ and $j$ that 
only goes through nodes $1, 2, \dots, k-1$, or 
2) paths between $i$ and $k$ and between $k$ and $j$ exist that only go through nodes $1, 2, \dots, k-1$. 
This observation leads to the following algorithm.  

 \algbegin Algorithm~1 (Undirected graph variant of Warshall's algorithm). 
 
 \algstep 1. [Initialize.] 
If $i=j$ or a link between $i$ and $j$ exists in $G$, then $r_{i,j}^{(0)} \leftarrow 1$; 
otherwise, $r_{i,j}^{(0)} \leftarrow 0$.
 
 \algstep 2. For $k = 1, 2, \dots, n$, repeat \textbf{a}: 
 
 \algstep ~~~a.~ For $(i, j)$ such that $1\leq i < j \leq n$,  
\[
r_{i,j}^{(k)} \longleftarrow r_{i,j}^{(k-1)} \ \lor \ (r_{i,k}^{(k - 1)} \ \land \ r_{k,j}^{(k - 1)} ). 
\]

Once the algorithm is executed, one of the following conditions holds. 


If nodes $i$ and $j$ are connected, $r_{i,j}^{(n)} = 1$. 

Otherwise, $r_{i,j}^{(n)} = 0$.




\section{Phase~1: Computing the maximum failure propagating stages}\label{sec:phase1}

Phase~1 computes the \emph{maximum failure propagation stage} 
among all possible choices of $k$ initial attack nodes. 
Let $l_{\max}(k)$ denote this number. 

The algorithm for this phase repeatedly solves  
SAT: the problem of determining whether a given Boolean formula 
is satisfiable, that is, whether there is a truth value assignment that makes the Boolean formula 
evaluate to true. 
The well-known Cook-Levin theorem states that SAT is NP-complete; however, modern SAT solvers 
are often capable of solving large formulas that arise from practical problems. 

By using a SAT solver, the algorithm determines whether the failure propagation continues until a given Stage $l$.  
To this end, each of the Boolean formulas, denoted $M_l$, is formulated so that the following property holds. 
\begin{quote}
$M_l$ is satisfiable if and only if some links fail in Stage~$l$ for some choice of $k$ attack nodes. 
\end{quote}
The algorithm is as follows.

 \algbegin Algorithm~2 (Computing $l_{max}$). 

 \algstep 1.  The satisfiability of $M_2$ and $M_3$ is checked. 
 If both are unsatisfiable, the output is 1.
 If $M_2$ is satisfiable and $M_3$ is unsatisfiable, the output is 2. 
 Otherwise, $l \longleftarrow 4$ and go to Step~2.
 
 \algstep 2. [Loop.]  
The satisfiability of $M_l$ is checked. 
If it is satisfiable, then 
 $l \longleftarrow l + 1$ and the loop is repeated.
Otherwise, go to Step~3. 

\algstep 3. Output $l-1$. ~\\


 

 
%
%


The algorithm repeatedly checks for the satisfiability of $M_l$, with $l$ being incremented. 
When $M_l$ is satisfiable, some links may fail in Stage~$l$. 
In this case, $l$ is incremented and the process is repeated. 
Otherwise, no link can fail in Stage~$l$ regardless of how the $k$ attack nodes are chosen.
Then, $l-1$ is the output. 
Even when no links in Network~B fail in Stage~2, some links in Network~A can fail in Stage~3; 
this exceptional case is addressed in the first step of the algorithm.

\subsection{Boolean formulas}

$M_l$ is built using Boolean operators and Boolean variables. 
Boolean operators include $\neg, \land, \lor$, and $\leftrightarrow$, 
which are negation, logical `and', logical `or', and logical equality, respectively. 
The Boolean variables are as follows. 
\begin{itemize}
    \item $x_{i,j}^{(s,k)}$ ($1\leq i < j \leq n, 0\leq k \leq n, s = 1, 3, \dots$) 
This variable serves as the same role as that of the variable $r_{i,j}^{(k)}$ in Warshall's algorithm
for Network~A.
Here, $s$ indicates the stage. 
For example, 
if $x_{i,j}^{(s,0)} = 1$, the link between $i$ and $j$ remains in Network~A at the end of Stage~$s$. 
Similarly, 
if $x_{i,j}^{(s,n)} = 1$, $i$ and $j$ are connected.  

    \item $y_{i,j}^{(s,k)}$ ($1\leq i < j \leq n, 0\leq k \leq n, s = 0, 2, \dots$): 
This variable is the equivalent of $x_{i,j}^{(s,k)}$ for Network~B.  
For technical convenience, $s = 0$, instead of $s=1$, indicates Stage~1.  

\item $z_i$ $(1\leq i \leq n)$ represents whether node $i$ is an attack node: 
 $z_i = 1$ if and only if node~$i$ is an attack node.

\end{itemize}
For simplicity, $x_{i,j}^{(s, k)}$ and $y_{i,j}^{(s, k)}$ are denoted by $x_{j,i}^{(s, k)}$ and $y_{j,i}^{(s, k)}$, respectively. 


$M_l$ comprises several sub-formulas over these variables, as follows: 
\begin{itemize}
    \item 
$S$ is used to represent the situation in Stage~1.  

\item
$AB_s$ and $BA_s$ represent the propagation of failures from Stage~$s-1$ to Stage~$s$.

\item
$P_l$ represents whether a link can fail during Stage~$l$. 

\end{itemize}

$M_l$ is built from these sub-formulas as follows.
\begin{equation}
    \begin{split}
        M_l := S  \land  AB_2  \land  BA_3  \land \dots \land  BA_{l-1}  \land  AB_l  \land  P_l \\
        \textrm{for\ } l = 2, 4, 6, \dots
    \end{split}
\end{equation}

\begin{equation}
    \begin{split}
        M_l := S  \land  AB_2  \land  BA_3  \land \dots \land  AB_{l-1}  \land  BA_l  \land  P_l \\
        \textrm{for\ } l = 3, 5, 7, \dots
    \end{split}
\end{equation}
%
The details of the construction of these sub-formulas will now be explained.

\subsection{Stage~1}

The initial stage (Stage~1) is represented by $S$. 
Specifically, $S$ represents $k$ attack nodes and the links remaining in Networks~A and~B. 

A link $\{i, j\}$ in Network~A remains intact in Stage~1 if and only if 
neither node $i$ nor node $j$ is an attack node. 
This is represented as 
\begin{equation}
x_{i,j}^{(1,0)} \leftrightarrow (\neg z_i \land \neg z_j) \quad \textrm{for\ } \{i,j\} \in E_A. 
\end{equation}
If no link between $i$ and $j$ exists in $E_A$, then $X_{i,j}^{(1,0)}$ should be 0. 
This is represented by the following Boolean formula:
\begin{equation}
\neg x_{i, j}^{(1, 0)} \quad \textrm{for\ }  \{i,j\} \in U - E_A, 
\end{equation}
where $U = \{ \{i, j\} \mid 1\leq i < j \leq n \}$.

This formulation applies to Network~B as follows:  
\begin{equation}
y_{i,j}^{(0,0)} \leftrightarrow (\neg z_i \land \neg z_j) 
\quad \textrm{for\ }  \{i,j\} \in E_B
\end{equation}
\begin{equation}
\neg y_{i, j}^{(0, 0)} 
\quad \textrm{for\ }  \{i,j\} \in U - E_B
\end{equation}

The number of attack nodes is $k$; any $k$ nodes can be attack nodes.
Hence, we need to enforce exactly $k$ $z_i$s to obtain the true value. 
\emph{Cardinality constraints}~\cite{10.1007/11564751_73} can be used for this purpose. 
A cardinality constraint is a Boolean formula that 
controls the number of variables that have the true value. 
We let $Card_{=k}(.)$ denote a cardinality constraint that evaluates to true 
if and only if exactly $k$ of the Boolean variables specified in the parentheses are assigned true. 


$S$ is a conjunction of all these sub-formulas. 
\begin{equation}
\begin{split}
S := & \bigwedge_{\{i, j\} \in E_A}  \Big( x_{i,j}^{(1,0)} \leftrightarrow (\neg z_i \land \neg z_j) \Big) \\ 
& \land \bigwedge_{\{i, j\} \in E_B} \Big(  y_{i,j}^{(0,0)} \leftrightarrow (\neg z_i \land \neg z_j) \Big) \\
& \land \bigwedge_{\{i,j\} \in U - E_A} \neg x_{i, j}^{(1, 0)} \land \bigwedge_{\{i,j\}\in U - E_B} \neg y_{i, j}^{(0, 0)} \\
& \land \ Card_{=k}(z_1, z_2,\dots,z_n)
\end{split}
\end{equation}

\subsection{Propagation of failures}

The Boolean formulas $AB_s$ and $BA_s$ are used to represent the failure propagation 
from Stage~$s-1$ to Stage~$s$. 
In even-numbered stages (Stages~$s = 2, 4, ...$), the failures in Network~A affects Network~B. 
$AB_s$ represents this propagation step, namely, the removal of links in Network~B during Stage~$s$. 
Similarly, when $s$ is odd ($s = 3, 5, ...$), $BA_s$ represents 
the removal of links in Network~A during Stage~$s$. 

To represent failure propagation, the connectivity between nodes must be 
expressed to determine the removal or non-removal of links. 
To this end, the formulas $AB_s$ and $BA_s$ encode Warshall's algorithm. 



In $AB_s$, the connectivity in Network~A at the end of Stage~$s-1$ must be expressed 
by the variables $x_{i,j}^{(s-1,k)}$. 
The Boolean formulas for this are derived directly from the algorithm, as follows. 
\begin{equation}\label{eq:x}
\begin{split}
    x_{i,j}^{(s-1, k)} \leftrightarrow 
    \Big( x_{i,j}^{(s-1, k-1)} \lor 
(x_{i,k}^{(s-1, k-1)} \land x_{k,j}^{(s-1, k-1)}) \Big) \\ 
\quad \textrm{for \ } k \in \{1, 2, \dots, n\} - \{i, j\} 
\end{split}
\end{equation}
As a result, 
$x_{i,j}^{(s-1,n)}$ indicates the existence of a path between $i$ and $j$ at the end of Stage~$s-1$.
Note that if $k=i$ or $k=j$,  $x_{i,i}^{(s-1, k-1)}$ or $x_{j,j}^{(s-1, k-1)}$ can be omitted. 
Hence, Formula~(\ref{eq:x}) is simplified as follows: 
\begin{equation}
    x_{i,j}^{(s-1, k)} \leftrightarrow x_{i,j}^{(s-1, k-1)}
\quad \textrm{for \ } k = i \textrm{ or } j
\end{equation}

Whether the link between $i$ and $j$ remains intact in Network~B at the end of Stage~$s$ 
depends on their connectivity in Network~A. 
This is represented in the form of the Boolean formula  
\begin{equation}
y_{i,j}^{(s,0)} \leftrightarrow \Big( x_{i,j}^{(s-1,n)} \land y_{i,j}^{(s-2, 0)} \Big) \quad 
\textrm{for \ } \{i,j\} \in E_B.
\end{equation}
Note that $y_{i,j}^{(s, 0)}$ and $y_{i,j}^{(s - 2, 0)}$ represent 
whether a link between $i$ and $j$ in Network~B exists at the end of Stages~$s$ and $s-2$, respectively. 

$y_{i,j}^{(s,0)}$ must always be false (0) if the link does not exist between $i$ and $j$ in the first place. 
This is represented as 
\begin{equation}
\neg y_{i,j}^{(s,0)}  \quad \textrm{for \ } \{i,j\} \in U -  E_B.
\end{equation}

Consequently, $AB_s$ is constructed as follows: 
\begin{equation}
\begin{split}
AB_s := & 
\bigwedge_{1\leq i<j \leq n} \Big( \bigwedge_{\genfrac{}{}{0pt}{3}{1\leq k \leq n}{k\neq i, j}} \Big( x_{i,j}^{(s-1, k)} \leftrightarrow \\
& \quad \big( x_{i,j}^{(s-1, k-1)} \lor (x_{i,k}^{(s-1, k-1)} \land x_{k,j}^{(s-1, k-1)}) \big) \Big) \\ 
& \quad \land (x_{i,j}^{(s-1, i)} \leftrightarrow x_{i,j}^{(s-1, i-1)}) \\
& \quad \land (x_{i,j}^{(s-1, j)} \leftrightarrow x_{i,j}^{(s-1, j-1)})
\Big) \\
&  \land \  \bigwedge_{\{i,j\}\in E_B} \Big(  y_{i,j}^{(s,0)} \leftrightarrow \big( x_{i,j}^{(s-1,n)}\land y_{i,j}^{(s-2, 0)} \big) \Big)  \\
&  \land \ \bigwedge_{\{i,j\}\in U - E_B} \neg y_{i,j}^{(s,0)}
\end{split}
\end{equation}

Failure propagation in odd-numbered stages ($s = 3, 5, ...$) is represented by 
the Boolean formula $BA_s$. 
This formula is constructed in the same manner as $AB_s$, except that 
$y_{i,j}^{(s,k)}$ and $x_{i,j}^{(s,k)}$ are swapped. 
\begin{equation}
\begin{split}
BA_s := & 
\bigwedge_{1\leq i<j \leq n} \Big( \bigwedge_{\genfrac{}{}{0pt}{2}{1\leq k \leq n}{k\neq i, j}} \Big( y_{i,j}^{(s-1, k)} \leftrightarrow \\
& \quad \big( y_{i,j}^{(s-1, k-1)} \lor (y_{i,k}^{(s-1, k-1)} \land y_{k,j}^{(s-1, k-1)}) \big) \Big) \\ 
& \quad \land (y_{i,j}^{(s-1, i)} \leftrightarrow y_{i,j}^{(s-1, i-1)}) \\
& \quad \land (y_{i,j}^{(s-1, j)} \leftrightarrow y_{i,j}^{(s-1, j-1)})
\Big) \\
&  \land \  \bigwedge_{\{i,j\}\in E_A} \Big(  x_{i,j}^{(s,0)} \leftrightarrow \big( y_{i,j}^{(s-1,n)}\land x_{i,j}^{(s-2, 0)} \big) \Big)  \\
&  \land \ \bigwedge_{\{i,j\}\in U - E_A} \neg x_{i,j}^{(s,0)}
\end{split}
\end{equation}

\subsection{Possibility of failures in Stage~$l$}

The Boolean formula $P_l$ is used to check whether link removal can occur during Stage~$l$, provided that any $k$ nodes can be attack nodes. 
It is sufficient to check whether there is at least one link that 
remained in Stage~$l-2$ but failed in Stage~$l$. 
When $l$ is even, we have 
\begin{equation}
P_l := \bigvee_{(i,j)\in E_B} (y_{i,j}^{(l-2,0)}\land \neg y_{i,j}^{(l,0)}).
\end{equation}
When $l$ is odd, we have
\begin{equation}
P_l := \bigvee_{(i,j)\in E_A} (x_{i,j}^{(l-2,0)}\land \neg x_{i,j}^{(l,0)}).
\end{equation}

\subsection{Example}

\begin{figure}
    \centering
    \includegraphics[scale=0.75]{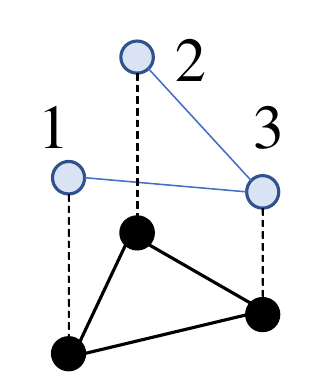}
    \caption{Three-node system.}
    \label{fig:smallexample}
\end{figure}

Consider the simple example shown in Fig.~\ref{fig:smallexample}  
and assume that $k = 1$. 
Algorithm~2 first checks whether $M_l$ is satisfiable with the SAT solver. 
The sub-formulas that compose $M_2 := S \land AB_2 \land P_2$ are as follows. 

\begin{equation}
\begin{split}
S := 
& \
\Big( x_{1,3}^{(0,0)} \leftrightarrow (\neg z_1 \land \neg z_3) \Big)  \\
& \
\land \Big( x_{2,3}^{(0,0)} \leftrightarrow (\neg z_2 \land \neg z_3) \Big) \\
& \
\land \neg x_{1,2}^{(0, 0)} \\
& \
\land \Big( y_{1,2}^{(1,0)} \leftrightarrow (\neg z_1 \land \neg z_2) \Big) \\
& \
\land \Big( y_{1,3}^{(1,0)} \leftrightarrow (\neg z_1 \land \neg z_3) \Big) \\
& \
\land \Big( y_{2,3}^{(1,0)} \leftrightarrow (\neg z_2 \land \neg z_3) \Big) \\
& \
\land (\neg z_1 \lor \neg z_2) \land (\neg z_1 \lor \neg z_3) \land (\neg z_2 \lor \neg z_3) 
 \\ 
& 
\quad \land (z_1 \lor z_2 \lor z_3)
\end{split}
\end{equation}
\begin{equation}
\begin{split}
AB_2 := 
& \ ( x_{1,2}^{(1, 1)} \leftrightarrow x_{1,2}^{(1, 0)} ) \\
& \land ( x_{1,2}^{(1, 2)} \leftrightarrow x_{1,2}^{(1, 1)} ) \\
& \land \Big( x_{1,2}^{(1, 3)} \leftrightarrow 
\big( x_{1,2}^{(1, 2)} \lor (x_{1,3}^{(1, 2)} \land x_{2,3}^{(1, 2)}) \big) \Big) \\
& \land ( x_{1,3}^{(1, 1)} \leftrightarrow x_{1,3}^{(1, 0)} ) \\
& \land \Big( x_{1,3}^{(1, 2)} \leftrightarrow 
\big( x_{1,3}^{(1, 1)} \lor (x_{1,2}^{(1, 1)} \land x_{2,3}^{(1, 1)}) \big) \Big) \\
& \land ( x_{1,3}^{(1, 3)} \leftrightarrow x_{1,3}^{(1, 2)} ) \\
& \land \Big( x_{2,3}^{(1, 1)} \leftrightarrow 
\big( x_{2,3}^{(1, 0)} \lor (x_{1,2}^{(1, 0)} \land x_{1,3}^{(1, 0)}) \big) \Big) \\
& \land ( x_{2,3}^{(1, 2)} \leftrightarrow x_{2,3}^{(1, 2)} ) \\
& \land ( x_{2,3}^{(1, 3)} \leftrightarrow x_{2,3}^{(1, 3)} ) \\
& \land \Big(  y_{1,2}^{(2,0)} \leftrightarrow \big( x_{1,2}^{(1,3)}\land y_{1,2}^{(0, 0)} \big) \Big) \\
& \land \Big(  y_{1,3}^{(2,0)} \leftrightarrow \big( x_{1,3}^{(1,3)}\land y_{1,3}^{(0, 0)} \big) \Big) \\
&  \land \Big(  y_{2,3}^{(2,0)} \leftrightarrow \big( x_{2,3}^{(1,3)}\land y_{2,3}^{(0, 0)} \big) \Big) \\ 
\end{split}
\end{equation}
\begin{equation}
\begin{split}
P_2 := 
& (y_{1,2}^{(0,0)} \land \neg y_{1,2}^{(2.0)}) \land (y_{1,3}^{(0,0)} \land \neg y_{1,3}^{(2.0)}) \\
& \land (y_{2,3}^{(0,0)} \land \neg y_{2,3}^{(2.0)}) 
\end{split}
\end{equation}
$M_2$ is satisfied by assigning 1 to $z_3$. 
This corresponds to the case where node~3 is an attack node. 

Next, the satisfiability of $M_3$ is checked.
Because $M_3$ is unsatisfiable, the algorithm terminates, and 2 is output.
In this case, $l_{max}$, the stage number of the last stage where a failure can occur, is two. 

\section{Phase~2: Finding the set of critical nodes}\label{sec:phase2}

Phase~2 solves the critical-node detection problem 
by finding the set of $k$ attack nodes that minimize the size of the largest 
connected component remaining at the end of Stage~$l_{max}$. 
To solve the minimization problem, ILP was ysed. 

The ILP formulation is similar to the SAT formulation for Phase 1, 
except that all variables are of integer type in the ILP, and the constraints 
are linear equations. 

\subsection{Variables}
As in Phase~1, we used variables 
$x_{i,j}^{(s,k)}$, $y_{i,j}^{(s,k)}$, and $z_i$; however, these are now 0-1 integer variables. 
In addition to these variables, we introduce a new variable, $bound$, 
which represents the upper bound on the size of 
connected components remaining after the cascading failure stopped. 
The variables used in the ILP formulation are summarized as follows: 
\begin{itemize}
    \item $x_{i,j}^{(s,k)}$ ($1\leq i < j \leq n, 0\leq k \leq n, s = 1, 3, \dots$):
0-1 variable representing the connectivity between nodes $i$ and $j$ in Network~A.  
At the end of Stage~$s$, the link between $i$ and $j$ exists  
if $x_{i,j}^{(s,0)} = 1$, 
and 
$i$ and $j$ are connected if $x_{i,j}^{(s,n)} = 1$. 

    \item $y_{i,j}^{(s,k)}$ ($1\leq i < j \leq n, 0\leq k \leq n, s = 0, 2, \dots$): 
0-1 variable, which is the equivalent of $x_{i,j}^{(s,k)}$ for Network~B. 

\item $z_i$ $(1\leq i \leq n)$: 0-1 variable representing an attack node. 
 $z_i = 1$ if and only if node~$i$ is an attack node.

\item $bound$: Integer variable representing an upper bound of the size of failure-induced connected components. 
\end{itemize}


\subsection{Linear constraints}

The constraints of the linear equation are mostly obtained by transforming 
the Boolean formulas used in Phase~1. 
For example, consider the following Boolean formula: 
\begin{equation}\label{eq:non-cnf}
    x_{i,j}^{(s-1, k)} \leftrightarrow \big( x_{i,j}^{(s-1, k-1)} \lor 
(x_{i,k}^{(s-1, k-1)} \land x_{k,j}^{(s-1, k-1)}) \big).
\end{equation}
As stated in Section~\ref{sec:phase1}, this formula applies to $AB_s$. 
Any Boolean formula can be transformed into a conjunctive normal form (CNF). 
A CNF formula is a conjunction of \emph{clauses}, each of which 
is a disjunction of variables or their negations. 
The CNF formula equivalent to Formula~(\ref{eq:non-cnf}) is   
\begin{equation}
\begin{array}{l}
 (\neg x_{i,j}^{(s-1, k)} \lor x_{i,j}^{(s-1, k-1)} \lor x_{i,k}^{(s-1, k-1)} ) \\
 \land \
 (\neg x_{i,j}^{(s-1, k)} \lor x_{i,j}^{(s-1, k-1)} \lor x_{k,j}^{(s-1, k-1)}) \\
 \land \
  (\neg x_{i,j}^{(s-1, k-1)} \lor x_{i,j}^{(s-1, k)})  \\
\land \ (\neg x_{i,k}^{(s-1, k-1)}  \lor \neg x_{k,j}^{(s-1, k-1)} \lor x_{i,j}^{(s-1, k)}).
\end{array}
\end{equation}
From this CNF Boolean formula, the following set of linear equations can be obtained: 
\begin{equation}
\begin{array}{l}
(1- x_{i,j}^{(s-1,k)}) + x_{i,j}^{(s-1,k-1)} + x_{i,k}^{(s-1,k-1)}  \geq 1, \\
(1- x_{i,j}^{(s-1,k)}) + x_{i,j}^{(s-1,k-1)} + x_{k,j}^{(s-1,k-1)}  \geq 1, \\
(1- x_{i,j}^{(s-1,k-1)}) + x_{i,j}^{(s-1,k)}  \geq 1, \\
(1- x_{i,k}^{(s-1,k-1)})  + (1- x_{k,j}^{(s-1,k-1)}) + x_{i,j}^{(s-1,k)} \geq 1. \\
\end{array}
\end{equation}



\subsubsection{Stage~1}

The linear constraints representing Stage~1 are obtained from Boolean formula $S$, 
except for the cardinality constraint, which represents $k$ attack nodes.
$S$ is transformed into the following linear equation: 

\begin{equation}
\left\{
\begin{array}{l}
1 \geq x_{i,j}^{(1,0)}+ z_i\\
1 \geq x_{i,j}^{(1,0)}+ z_j\\
x_{i,j}^{(1,0)} + z_i + z_j \geq 1
\end{array}
\right.
\quad \textrm{for\ } \{i,j\} \in E_A, 
\end{equation}
\begin{equation}
x_{i, j}^{(1, 0)} = 0 \qquad \textrm{for\ } \{i,j\} \in U - E_A, 
\end{equation}

\begin{equation}
\left\{
\begin{array}{l}
1 \geq y_{i,j}^{(0,0)}+ z_i\\
1 \geq y_{i,j}^{(0,0)}+ z_j\\
y_{i,j}^{(0,0)} + z_i + z_j \geq 1
\end{array}
\right.
\quad \textrm{for\ } \{i,j\} \in E_B,
\end{equation}
\begin{equation}
y_{i, j}^{(0, 0)} = 0 \qquad \textrm{for\ } \{i,j\} \in U - E_B.
\end{equation}

The cardinality constraint, which represents that any $k$ nodes can be used as attack nodes,  
is replaced by the following equation: 
\begin{equation}
\sum_{i=1}^n z_i = k. 
\end{equation}


\subsubsection{Propagation of failures}

Linear constraints representing stage-by-stage 
failure propagation are obtained from the Boolean formulas $AB_l$ and $BA_l$ 
in the same manner as described above. 

The linear constraints shown below correspond to the Boolean formulas $AB_2, AB_4, \ldots$. 
\begin{equation}
\label{eq:const1}
\begin{split}
\left\{
\begin{array}{l}
(1- x_{i,j}^{(s-1,k)}) + x_{i,j}^{(s-1,k-1)} + x_{i,k}^{(s-1,k-1)} \geq 1 \\
(1- x_{i,j}^{(s-1,k)}) + x_{i,j}^{(s-1,k-1)} + x_{k,j}^{(s-1,k-1)} \geq 1 \\
(1- x_{i,j}^{(s-1,k-1)}) + x_{i,j}^{(s-1,k)} \geq 1 \\
(1- x_{i,k}^{(s-1,k-1)})  + (1- x_{k,j}^{(s-1,k-1)}) + x_{i,j}^{(s-1,k)} \geq 1 \\
\end{array}
    \right.\\
\textrm{for \ } 1\leq i, j \leq n, k \in \{1, 2, \dots, n\} - \{i, j\}, \\
s = 2, 4, \dots, 2\lceil  l_{max}/2 \rceil
\end{split}
\end{equation}

\begin{equation}
\label{eq:const2}
\begin{split}
x_{i,j}^{(s-1,k-1)} =  x_{i,j}^{(s-1,k)}  \\
\textrm{for \ } 1\leq i, j \leq n, k \in \{i, j\}, s = 2, 4, \dots, 2\lceil l_{max}/2 \rceil
\end{split}
\end{equation}

\begin{equation}
\begin{split}
\left\{
\begin{array}{l}
(1 - y_{i,j}^{(s,0)}) + x_{i,j}^{(s-1, n)} \geq 1  \\
(1 - y_{i,j}^{(s,0)}) + y_{i,j}^{(s-2, 0)} \geq 1  \\
(1 - x_{i,j}^{(s-1,n)}) + (1 - y_{i,j}^{(s-2, 0)}) + y_{i,j}^{(s,0)} \geq 1  \\
\end{array}
    \right.\\
\textrm{for \ } \{i, j\} \in E_B,  s = 2, 4, \dots, 2\lfloor l_{max}/2\rfloor
\end{split}
\end{equation}

\begin{equation}
\begin{split}
y_{i,j}^{(s,0)} = 0  \\
\textrm{for \ } \{i, j\} \in U - E_B, s = 2, 4, \dots, 2\lfloor l_{max}/2 \rfloor
\end{split}
\end{equation}

A slight difference from the SAT formulation is that 
when $l_{max}$ is odd, the constraints (\ref{eq:const1}) and (\ref{eq:const2}) 
are necessary for Stage~$s = l_{max} + 1$. 
Although further failures never occur in Stage~$l_{max}+1$ or later, 
this is necessary to express the size of failure-induced connected components. 

The linear constraints shown below correspond to the Boolean formulas $BA_3, BA_5, \ldots$. 
\begin{equation}\label{const3}
\begin{split}
\left\{
\begin{array}{l}
(1- y_{i,j}^{(s-1,k)}) + y_{i,j}^{(s-1,k-1)} + y_{i,k}^{(s-1,k-1)} \geq 1 \\
(1- y_{i,j}^{(s-1,k)}) + y_{i,j}^{(s-1,k-1)} + y_{k,j}^{(s-1,k-1)} \geq 1 \\
(1- y_{i,j}^{(s-1,k-1)}) + y_{i,j}^{(s-1,k)} \geq 1 \\
(1- y_{i,k}^{(s-1,k-1)})  + (1- y_{k,j}^{(s-1,k-1)}) + y_{i,j}^{(s-1,k)} \geq 1 \\
\end{array}
    \right.\\
\textrm{for \ } k \in \{1, 2, \dots, n\} - \{i, j\}, \\
s = 3, 5, \dots, 2\lceil \tfrac{l_{max}-1}{2} \rceil+1
\end{split}
\end{equation}

\begin{equation}\label{const4}
\begin{split}
y_{i,j}^{(s-1,k-1)} =  y_{i,j}^{(s-1,k)}  \\
\textrm{for \ } k \in \{i, j\}, s = 3, 5, \dots, 2\lceil \tfrac{l_{max}-1}{2} \rceil+1
\end{split}
\end{equation}

\begin{equation}
\begin{split}
\left\{
\begin{array}{l}
(1 - x_{i,j}^{(s,0)}) + y_{i,j}^{(s-1, n)} \geq 1  \\
(1 - x_{i,j}^{(s,0)}) + x_{i,j}^{(s-2, 0)} \geq 1  \\
(1 - y_{i,j}^{(s-1,n)}) + (1 - x_{i,j}^{(s-2, 0)}) + x_{i,j}^{(s,0)} \geq 1  \\
\end{array}
    \right.\\
\textrm{for \ } k \in \{1, 2, \dots, n\} - \{i, j\}, \\
s = 3, 5, \dots, 2\lfloor \tfrac{l_{max}-1}{2} \rfloor+1
\end{split}
\end{equation}

\begin{equation}
\begin{split}
x_{i,j}^{(s,0)} = 0  \\
\textrm{for \ } \{i, j\} \in U - E_A, s = 3, 5, \dots, 
2\lfloor \tfrac{l_{max}-1}{2} \rfloor+1
\end{split}
\end{equation}


Similar to the constraints (\ref{eq:const1}) and (\ref{eq:const2}), 
the constraints (\ref{const3}) and (\ref{const4}) are necessary for 
Stage~$s = l_{max} + 1$ if $l_{max}$ is even. 

\subsubsection{Connected component size}

To formulate the objective function, 
we require some linear constraints to represent the size of the maximum connected component at the end of Stage~$l_{max}$.

Consider the case in which $l_{max}$ is odd. 
Suppose that $ x_{i,j}^{(l_{max},n)} \ (1 \leq i < j \leq n) $ are set to their intended values, 
that is, $x_{i,j}^{(l_{max},n)} = 1$ if nodes~$i$ and $j$ are connected at the end of Stage~$l_{max}$, 
otherwise $x_{i,j}^{(l_{max},n)} = 0$. 
In this case, the size of the failure-induced connected component to which node $i$ belongs is one plus
the summation of $x_{i,j}^{(l_{max},n)}$ for all $j$. 
To indirectly represent the largest size among the failure-induced connected components, the constraints use the integer variable $bound$, representing the upper bound of the sizes. 
When $l_{max}$ is odd, the constraints are as follows:   
\begin{equation}
\begin{split}
        \sum_{1\leq j < i, i <j \leq n} x_{i,j}^{(l_{max},n)} < bound  \\
        \quad \textrm{for\ } 1\leq i \leq n.
\end{split}
\end{equation}


\noindent
When $l_{max}$ is even, the constraints are as follows:
\begin{equation}
\begin{split}
        \sum_{1\leq j < i, i <j \leq n} y_{i,j}^{(l_{max},n)} < bound  \\
        \quad \textrm{for\ } 1\leq i \leq n.
\end{split}
\end{equation}

\subsection{Objective function}
The objective function is expressed as follows:
\begin{quote}
Minimize \qquad $bound$ 
\end{quote}
By minimizing $bound$, the size of the maximum failure-induced connected component is minimized. 
The values of variable $z_i$ of the optimal solution correspond to the 
critical nodes;  
the set of critical nodes is $\{i \mid z_i = 1\}$. 

\section{Experiments}\label{sec:experiment}

\subsection{Experiment settings}
We conducted experiments to answer the following questions. 
\begin{itemize}
    \item How large are the problems that the proposed approach can solve?
    \item How good is the optimal solution compared to non-optimal solutions obtained by heuristic approaches?
\end{itemize}
The second question is important because the critical node detection problem is NP-hard, and, thus 
we must to resort to non-optimal approaches for large problem instances. 

We implemented the proposed approach using Python and the well-known SAT and ILP solvers 
Z3~\cite{z3} and the Gurobi optimizer~\cite{gurobi} 
(technically, Z3 is called an SMT solver because it can solve the satisfability problem 
for more expressive logic than Boolean algebra.)
All experiments were conducted using a Windows~10 PC laptop equipped with 
an Intel Core i7-7700HQ CPU (3.8GHz) and 16GB of memory. 


The problem instances used in the experiments were synthesized as follows. 
For each instance, the initial topologies of the power network (Network~A) and communication network (Network~B) 
were first constructed. 
This was performed based on the extended Barabasi-Albert model~\cite{barabasi2000}, which is a well-known scale-free network model. 
In a scale-free network, the degree of the nodes follows a power-law distribution, that is, 
$P(d) \propto d^{-\beta}$, where 
$P(d)$ is the fraction of nodes with $d$ links, and $\beta$ is a parameter. 
We set the parameter $\beta$ to 3.0 for the power network and 2.2 for the communication network because these numbers were observed in 
real-world networks~\cite{barabasi1999}. 
To create these networks, we used the NetworkX Python package for studying complex networks~\cite{SciPyProceedings_11}.
If the created network is disconnected, then 
it is discarded and a new one is created again. 

Next, a one-to-one map between the sets of nodes of the two networks was generated to match the nodes of the two networks. 
For this task, we used the CAS algorithm, which was used to synthesize interdependent networks in~\cite{Nguyen2013}. 
This algorithm is probabilistic but associates high-degree nodes in one network with those in the other, and 
associates low-degree nodes with low-degree nodes. 

We consider two non-exact heuristics algorithms. 
The first, which we simply call the \emph{greedy heuristic algorithm}  (Algorithm~3), 
selects attack nodes individually until $k$ nodes are chosen. 
A node is selected as a new attack node if the maximum failure-induced connected component is minimized when, in addition to the already-selected attack nodes, that node is an attack node. 
If there is more than one such node, 
the node with the greatest summation of degrees, that is, 
the node that has the greatest number of incident links, is chosen.  
 
 \algbegin Algorithm~3 (Greedy heuristic algorithm). 
 
 \algstep 1. [Initialize.] 
 Set $K \longleftarrow \emptyset$, 
 $G_1 \longleftarrow G_A$, $G_2 \longleftarrow G_B$, and 
 $V \longleftarrow \{1,2,\cdots,n\}$. 
 
 \algstep 2. Repeat $k$ times: 

 \algstep ~~~a.~ For all $i \in V$, do: 

 \algstep ~~~~i.~  Simulate failure propagation provided that $K\cup \{i\}$ is the set of attack nodes. 
 
 \algstep ~~~~ii.~ Set $s_i$ to be the size of the maximum failure-induced connected component. 

 \algstep ~~~b.~ 
 Set $C \longleftarrow \{j \in V \mid s_j = \min_{i\in V} \{s_i\} \}$.

 \algstep ~~~c.~ 
 One node $i_{atk}$ is selected from $C$ such that $d(i_{atk}) = \max_{i\in C}\{d(i)\}$, 
 where $d(i)$ denotes the summation of $i$'s degree for $G_1$ and $G_2$.  

 \algstep ~~~d.~ 
Set $K \longleftarrow K \cup \{i_{atk}\}$. 
 Remove the node $i_{atk}$ and its incident links from $G_1$ and $G_2$.
 Set $V \longleftarrow V - \{i_{atk}\}$.~\\

The second algorithm is an adaptation of the Max-Cas heuristic algorithm~\cite{Nguyen2013}. 
This algorithm was originally proposed for the simplified model (see~Section \ref{sec:related}) but can  readily be adapted  
for the interdependent network model assumed in this study. 
Similar to the greedy heuristic algorithm, the algorithm chooses the attack nodes one by one,  
but \emph{articulation nodes} are first chosen as candidates for an attack node. 
An articulation node is a node whose removal immediately splits the graph. 
If no articulation nodes exist in either network, all nodes are chosen as candidates. 
In a sense, the Max-Cas algorithm is a run-time optimized version of the greedy heuristic algorithm because it can avoid failure simulations for non-articulation nodes. 
The original Max-Cas algorithm does not define the selection of one node 
when two or more nodes satisfy the condition of being attack nodes. 
The adapted version selects the one with the most incident links, just as  
the greedy heuristic algorithm does. 


\algbegin Algorithm~4 (Modified Max-Cas heuristic algorithm). 
 
 \algstep 1. [Initialize.] 
 Set $K \longleftarrow \emptyset$.  
 Set $G_1 \longleftarrow G_A$, $G_2 \longleftarrow G_B$.
 Set $V \longleftarrow \{1,2,\cdots,n\}$. 
 
 \algstep 2. Repeat $k$ times: 
 
 \algstep ~~~a.~  Set $A$ to be the set of articulation nodes of graphs $G_1$ and $G_2$. 

 \algstep ~~~b.~  If $A=\emptyset$, set $C \longleftarrow V$ and go to \textbf{e}.

\algstep ~~~c.~ For all $i \in A$, do:

 \algstep ~~~~i.~  Simulate failure propagation provided that $K\cup \{i\}$ is the set of attack nodes. 
 
 \algstep ~~~~ii.~ Set $s_i$ to be the size of the maximum failure-induced connected component. 

 \algstep ~~~d.~ 
 Set $C \longleftarrow \{j \in A \mid s_j = \min_{i\in A} \{s_i\} \}$.

 \algstep ~~~e.~ 
 One node $i_{atk}$ is selected from $C$ such that $d(i_{atk}) = \max_{i\in C}\{d(i)\}$, 
 where $d(i)$ denotes the summation of $i$'s degree for $G_1$ and $G_2$.  

 \algstep ~~~d.~ 
Set $K \longleftarrow K \cup \{i_{atk}\}$. 
 Remove the node $i_{atk}$ and its incident links from $G_1$ and $G_2$.
 Set $V \longleftarrow V - \{i_{atk}\}$.~\\

 
 
 

 
 


\subsection{Results}

We created a collection of problem instances with 20, 25, and 30 nodes and 
applied the proposed approach and two heuristic algorithms to each of them.  
We changed the number of instances created depending on the number of nodes. 
Specifically, we created 121 instances for 20 nodes, 21 instances for 25 nodes, 
and 11 instances for 30 nodes. 
This is because, as shown below, the proposed approach required a much longer computation time for larger instances,
and we opted to obtain as many experimental results as possible within the limited period of the project. 
We set $k$, the number of attack nodes, to one-fifth the number of nodes. 

Table~\ref{tab:result1} lists the experimental results. 
For each number of nodes, we selected three cases where the running time of our approach was the shortest, median, and 
longest of all the problem instances of that number of nodes. 
Table~\ref{tab:timeave} summarizes the statistics related to the running time, and 
shows the average and variance for each of number of nodes. 
These results show that the running times are distributed very diversely. 
The longest time was three orders of magnitude longer than the shortest time for all three numbers of nodes. 
The time even reached approximately nine hours for one of the 30-node instance.
This indicates that the proposed approach can handle problems with up to 30~nodes using an ordinary PC. 

The breakup of the total running time shows that Phase~2, the ILP, caused a large diversity in the running time;  
it consumes more than 90 percent of the total time for the longest time cases, whereas only half or a smaller fraction of 
total time for the shortest time cases. 
The running time of Phase~1 also exhibited a large variance; however,  
this was significantly smaller than that in Phase~2.

The running time of the non-exact heuristic algorithms was very short compared to that of the proposed approach. 
The greedy heuristic algorithm usually requires an order of magnitude longer running time than the Max-Cas algorithm. 
However, it was still very short: much less than one second for all cases, including cases not shown in Table~\ref{tab:result1}. 

\begin{table*}
    \centering
        \caption{Experimental results.} 
    \label{tab:result1}
    \begin{tabular}{ccccccccccc}
    \hline
        &     &           \multicolumn{5}{c}{Proposed} 
        & \multicolumn{2}{c}{Greedy} & \multicolumn{2}{c}{Max-Cas} \\ 
    $n$ & $k$ & $l_{max}$ & $f$ & Ph.~1 & Ph.~2 & total & $f$ & time & $f$ & time \\ \hline
    20 & 4 & 3 & 5 & 4 & 6 & 10 & 5 & 0.0678 & 5 & 0.00499 \\
        20 & 4  &  5 & 5  & 10  & 73  & 83 & 5 & 0.06098 & 5 & 0.00597  \\
        20 & 4 &  7 &  5 & 22  & 7108  &  7130 & 9 & 0.06587 & 9 & 0.00553 \\
        25 & 5 & 6 & 2 & 47 & 33 & 80 & 2 & 0.10272 & 4 & 0.01297 \\
        25 & 5 & 7  & 4  &  95 & 1240  &  1335 & 5 & 0.08969 & 9 & 0.00646  \\
        25 & 5 & 6  & 5  & 46  & 8456  &  8502 & 11 & 0.15658 & 11 & 0.01134 \\
        30 & 6 & 10 & 2 & 656  & 47 & 703 & 2 & 0.18794 & 2 & 0.01978 \\
        30 & 6 &  8 & 2  & 676  & 1073  &  1749 & 13 & 0.20189 & 7 & 0.01452 \\
        30 & 6 &  12  & 2  & 2393   & 29000  &  31393 & 7 & 0.20004 & 9 & 0.00796 \\
        \hline
        \multicolumn{11}{l}{Three cases are shown for each size of problem instances.} \\
        \multicolumn{11}{l}{$n$: number of nodes. $k$: number of attack nodes.} \\
        \multicolumn{11}{l}{$f$: size of largest failure-induced connected component.} \\
        \multicolumn{11}{l}{Others: running times in seconds.} \\
    \end{tabular}
\end{table*}

\begin{sidewaystable}
    \centering
        \caption{Running time statistics. }
    \label{tab:timeave}
\begin{tabular}{lllllllllll}
\hline
   & \multicolumn{2}{c}{Proposed-Ph.1}                        & \multicolumn{2}{c}{Proposed-Ph.2}                        & \multicolumn{2}{c}{Proposed-total}                         & \multicolumn{2}{c}{Greedy}                              & \multicolumn{2}{c}{Max-Cas}                              \\
$n$  & \multicolumn{1}{c}{ave} & \multicolumn{1}{c}{var} & \multicolumn{1}{c}{ave} & \multicolumn{1}{c}{var} & \multicolumn{1}{c}{ave} & \multicolumn{1}{c}{var} & \multicolumn{1}{c}{ave} & \multicolumn{1}{c}{var} & \multicolumn{1}{c}{ave} & \multicolumn{1}{c}{var} \\
\hline
20 & 1.30E+01                & 6.92E+01                & 2.58E+02                & 6.91E+05                & 2.71E+02                & 6.97E+05                & 6.92E-02                & 7.66E-05                & 6.02E-03                & 9.35E-06                \\
25 & 6.36E+01                & 9.41E+02                & 2.60E+03                & 7.38E+06                & 2.67E+03                & 7.40E+06                & 1.11E-01                & 4.48E-04                & 1.01E-02                & 1.46E-05                \\
30 & 8.59E+02                & 4.35E+05                & 6.05E+03                & 9.08E+07                & 6.91E+03                & 9.84E+07                & 2.16E-01                & 6.49E-04                & 1.31E-02                & 3.52E-05         \\ \hline
\end{tabular}
\end{sidewaystable}

To determine the difference between the optimal solutions and those obtained by the heuristic algorithms, 
we compare them in terms of the size of the largest failure-induced connected component in Table~\ref{tb:size}. 
For each of the cases and each of the three algorithms, we computed the the size of the largest failure-induced connected component provided that the $k~(=n/5)$ nodes output by the algorithm were attack nodes. 
The table shows the average and variance of these sizes. 
The results show that the largest connected component size achieved by the heuristic algorithms usually remains within two times the optimal size on average, but has a wide distribution. 

\begin{table}
\centering
\caption{Largest failure-induced connected component size statistics.}
\label{tb:size}
\begin{tabular}{lllllll}
\hline 
\multicolumn{1}{c}{}  & \multicolumn{2}{c}{Proposed}                      & \multicolumn{2}{c}{Greedy}                        & \multicolumn{2}{c}{Max-Cas}                       \\ \hline
\multicolumn{1}{c}{n} & \multicolumn{1}{c}{ave} & \multicolumn{1}{c}{var} & \multicolumn{1}{c}{ave} & \multicolumn{1}{c}{var} & \multicolumn{1}{c}{ave} & \multicolumn{1}{c}{var} \\
20                    & 5.06                    & 4.45                    & 7.58                    & 11.86                   & 8.12                    & 12.04                   \\
25                    & 5.38                    & 6.05                    & 8.62                    & 19.19                   & 9.76                    & 18.75                   \\
30                    & 3.18                    & 3.06                    & 7.45                    & 24.43                   & 8.00                    & 18.18                  \\ \hline
\end{tabular}
\end{table}

\begin{figure}
    \centering
    \includegraphics[width=1\columnwidth,bb=0.000000 0.000000 504.000000 585.000000]{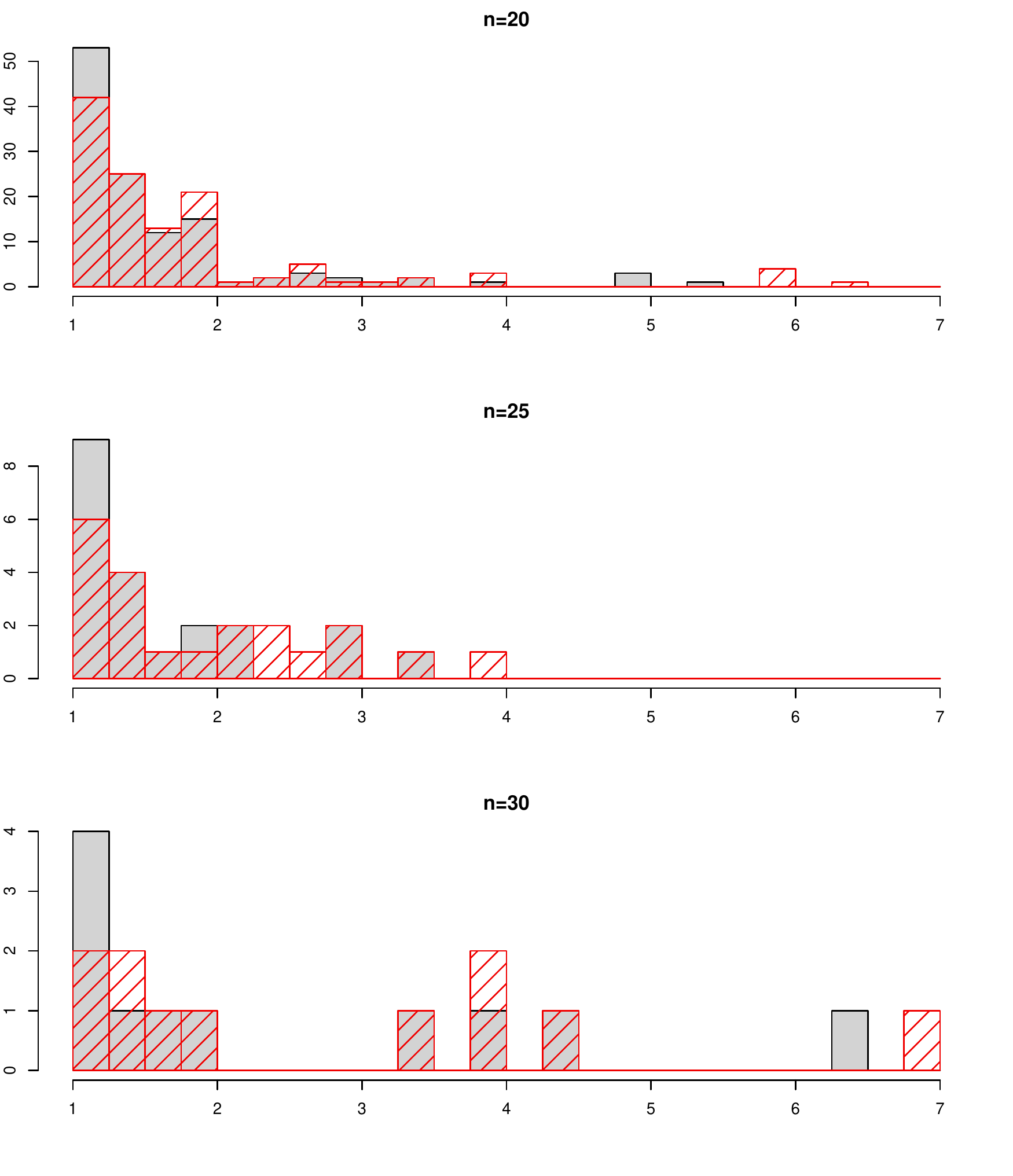}
    \caption{Histograms of the relative size of the largest failure-induced connected component obtained by the heuristic algorithms. 
    The grey bars and those filled with slanted lines correspond to the greedy  and Max-Cas heuristic algorithms, respectively.
    Each bin size is 0.25.  
    For example, the leftmost bin shows the cases ranging from 1.0 (in which case the optimal solution was obtained) to 1.25. }
    \label{fig:hist}
\end{figure}

The histograms in Fig.~\ref{fig:hist} show how close to or how far from the optimal solutions the heuristic solutions are. 
The histograms display the ratio of the optimal result yielded by the proposed approach to those obtained by the heuristic algorithms in terms of the size of the largest failure-induced connected components.  
For example, if the largest failure-induced connect component was of size 2 and 5 when the attack nodes were selected, respectively, by the proposed approach and one of the heuristic algorithms, then the ratio would be 2.5. 
The histograms show that the heuristic algorithms sometimes yield the optimal or near-optimal solutions, 
which correspond to the cases falling in the leftmost bin. 
At the same time, the heuristic algorithms often select attack nodes that result in a much larger connected component. 
This suggests that there is much room for improvement in the heuristic algorithms. 
It should be noted that this analysis is only possible because of the optimal solution to the critical node detection problem obtained using our proposed approach. 

\section{Conclusions}\label{sec:conclusion}

In this study, we addressed the critical node detection problem for interdependent networks.  
The proposed approach can solve this problem using SAT and ILP solvers,  
by encoding a stage-by-stage failure propagation using Boolean and integer linear constraints. 
We implemented this and two non-exact heuristic approaches in 
experiments, which showed that, although the problem is NP-hard, the proposed approach is capable of finding the optimal solution for problems of moderate size. 
The experimental results also showed that the two heuristic approaches were sometimes successful in finding an optimal or near optimal solution but often chose a node set that only causes benign effects. 

There are several directions for future research in this area. 
First, reducing the execution time of the proposed approach is of interest. 
This could be achieved, for example, by devising new SAT or ILP formulations. 
Another direction is to adapt the proposed approach to different models of interdependent networks. 
The development of new heuristic approaches also deserves further study.


\end{document}